# A Primer on Spreadsheet Analytics

Thomas A. Grossman

University of San Francisco, School of Business & Management, San Francisco CA 94117
tagrossman@usfca.edu

**ABSTRACT**

This paper provides guidance to an analyst who wants to extract insight from a spreadsheet model. It discusses the terminology of spreadsheet analytics, how to prepare a spreadsheet model for analysis, and a hierarchy of analytical techniques. These techniques include sensitivity analysis, tornado charts, and backsolving (or goal-seeking). This paper presents native-Excel approaches for automating these techniques, and discusses add-ins that are even more efficient. Spreadsheet optimization and spreadsheet Monte Carlo simulation are briefly discussed. The paper concludes by calling for empirical research, and describing desired features spreadsheet sensitivity analysis and spreadsheet optimization add-ins.

## 1. INTRODUCTION

Spreadsheets are widely used for modeling and analysis but there is little support for someone who needs to use a spreadsheet model to generate analytical insight. To provide guidance to a spreadsheet analyst, this paper presents an overview of spreadsheet analytics, including terminology, spreadsheet model preparation, and a hierarchy of analytical techniques. I describe how to implement the techniques in native Excel and discuss add-ins for automating them. The spreadsheet analytic concepts discussed include:

<u>Terminology</u>: model, input, output, data, decision variable, parameter, performance measure, case, scenario, strategy;

<u>Preparing a model for analysis</u>: accuracy, protection, master copy, design for analysis, base case selection, record base case, program change-from-base;

<u>Sensitivity analysis</u>: simple what-if analysis, one-parameter sensitivity analysis, tornado chart, two-parameter sensitivity analysis, multi-parameter sensitivity analysis (scenario analysis);

<u>Backsolving:</u> finding input value(s) that generate a desired output value. This is called "breakeven analysis" in accounting, "inverse analysis" in engineering, and "goal-seeking" in Excel);

I believe these to be the fundamental techniques for analyzing a model. I also discuss briefly advanced techniques for "what's best analysis" (optimization) and risk analysis (Monte Carlo simulation) that require add-ins for meaningful application. (This paper does not consider spider charts and decision trees because they are specialist tools used in the discipline of decision analysis.)

### 1.1. Literature Survey and Contribution

[Jennings 2003] discussed at a high level analytical techniques including sensitivity testing, scenario analysis, Monte Carlo simulation, optimization, and the emerging technique of real options. [Powell and Baker 2007] provide in chapter 6 a hierarchy of analytical techniques, including base-case, benchmarking (program change-from-base), sensitivity analysis and scenario analysis, breakeven analysis, optimization analysis, and risk analysis using Monte

Carlo simulation. [Read and Batson 1999] provide in chapter 8 some analytical tools with special attention to using a model over time, including Data\Table…\, reconciliation tables (used to manage model runs and distinguish between changes to input values and changes to spreadsheet code), and run trees (large scenarios).

This paper integrates and extends the ideas from previous work. I provide a more complete hierarchy of techniques, a unified nomenclature, and a clear distinction between a technique (which is applicable to all models—not only models coded in a spreadsheet) and its best spreadsheet implementation. I provide an improved scenario tool, and an overview of both Excel-native methods and software add-ins where applicable.

This paper is focused on forward-looking planning models of the type commonly used in business operations, finance, managerial accounting, consulting, investment banking etc. The ideas and techniques might be applicable to other classes of spreadsheet models.

### 1.2. Structure of Paper

The paper begins (section 2) by proposing terminology that is useful for talking about models and analysis. Section 3 discusses how to prepare an existing spreadsheet model to make it suitable for the application of analytical techniques. Section 4 presents the techniques of sensitivity analysis. Section 5 discusses backsolving (called "goal seek" in Excel). Sections 6 and 7 briefly discuss the add-in based techniques of spreadsheet optimization, and spreadsheet Monte Carlo simulation. Section 8 discusses the need for empirical research as well as desired enhancements to analytical spreadsheet add-ins.

### 2. TERMINOLOGY FOR ANALYSIS OF MODELS

There is no standard terminology for model analysis. I propose a set of terms here. A **model** is a set of mathematical equations. A model can be represented as algebraic formulae, as a spreadsheet, or as a procedural computer program. The terms below apply to models and model analysis independent of how the model is represented.

A numeric value required by a model is an **input**. A numeric value computed by a model is an **output**. One can think of a model as a machine for converting *inputs* to *outputs*. (In a spreadsheet, any cell containing a number is an *input* and any cell containing a formula is an *output*.) A particularly important output is called a **performance measure**. See Figure 1.

Many model analyses evaluate the impact of alternative decisions and seek to discover good or optimal decisions. Therefore it is useful to distinguish between two types of *inputs*. A **data** is a number whose value is outside one's control (i.e., a "fact"). A **decision variable** is a number whose value one can control (i.e., a "choice"). Any *input* to a model is designated by the analyst as a *data* or a *decision variable*. This designation can change with the timeframe of an analysis. For example, in a production planning model the production capacity might be *data* in a short-term analysis but a *decision variable* in a long-term analysis.

A set of values for all *data* is a **scenario**. A set of values for all *decision variables* is a **plan**. A set of values for all *inputs* (*data* and *decision variables*) is a **case**. Thus, a *case* comprises a *scenario* and a *plan*. The **base case** is the initial set of *input* values for a model. (Note that a model represented as a spreadsheet can be easier to build if it has a fully-specified base case at the time of programming; a model represented as algebraic formulae or procedural computer program does not need *base case* values until computation time.) The **base values** are all of the numeric values of a model (*inputs* and *outputs*) when the *inputs* are set to the *base case*. (In spreadsheet terms, the base values are all the numbers visible in the spreadsheet when the *inputs* are set to the *base case*.)

When performing sensitivity analysis, it is convenient for the analyst to designate one or more *inputs* for special attention. A **parameter** is an *input* that is systematically varied through a range of values. When a *decision variable* is parameterized, it teaches us about the impacts of different decisions. When a *data* is parameterized, it teaches us about the impact of different possible futures; the centerpiece of risk analysis is assumptions about *data* whose values we don't yet know. See Figure 1. (Note that "*parameter*" is sometimes used in the sense of "perimeter" or bounds of an analysis, or in the sense of "*inputs*".)

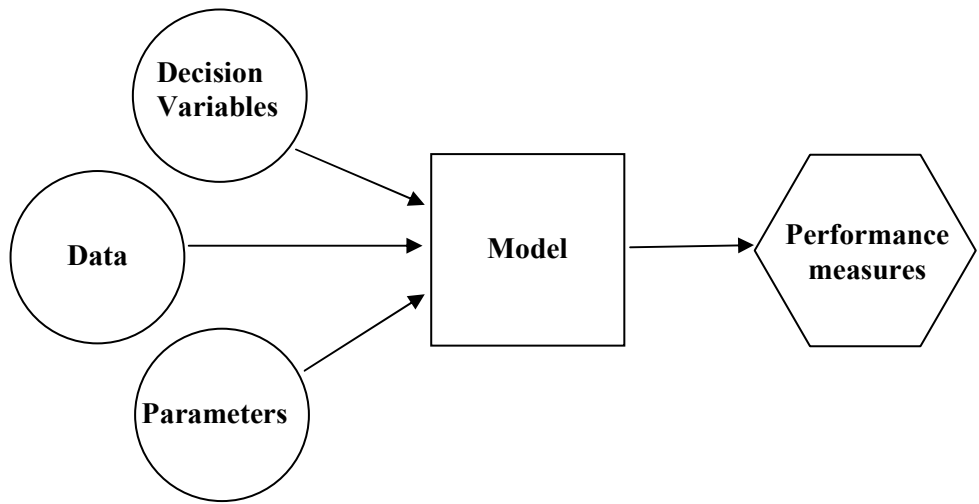

Figure 1: Relationship between Inputs (Decision Variables, Data, Parameters), Model, and Important Outputs (Performance Measures)

## 3. PREPARING A MODEL FOR ANALYSIS

The recommendations in section 3 are guidelines. An experienced analyst will sometimes choose to deviate from the guidelines. I recommend that analysts be able to clearly articulate the reasons when they choose not to follow the guidelines.

### 3.1. Preliminaries: Accuracy, Protection and Master Copy

The purpose of analysis is to obtain insight from a model. Therefore, the model should be correct (i.e., free of conceptual errors). The computer program that implements the model should be well-engineered and accurate. At the conclusion of programming (before performing analysis), protect all formula cells so they cannot inadvertently be changed. To insure that the model is not corrupted during analysis, make a "master copy". This can be done using a source code management system such as Microsoft SharePoint, or simply by making the spreadsheet read-only and storing the master copy in a special location.

### 3.2. Design for Analysis

There are many ways to design a spreadsheet, and there might be many different "good" spreadsheet designs for the same model. When a spreadsheet is intended for analysis it should be designed to facilitate efficient and accurate execution of analytical techniques.

There are several spreadsheet design-for-analysis principles. All inputs should be entered once in a single cell and then referenced as needed. Because it is difficult to keep track of inputs that are scattered throughout a spreadsheet, place all inputs in a single module or set of modules and "echo" them to cells where they are needed. It is helpful to separate data and decision variable inputs into separate sub-modules. Locate important inputs and important outputs so they can be seen together on a single screen without scrolling. When working with large models the analyst should create a "control panel" worksheet that houses key inputs and echoes key outputs.

It can be necessary to modify the design of an existing spreadsheet to facilitate analysis. Any redesign work should be carefully planned and checked to avoid the introduction of errors.

### 3.3. Base Case Selection

The *base case* (initial values for data and decision variables) is selected by the analyst. It should contain "typical" or "reasonable" values. They are chosen by exercising analytical and managerial judgment, guided by any relevant historical data. Because the analytical techniques described below allow the analyst to investigate any deviations from the base case, the base case does not need to be "perfect".

Be aware that a common technique employed by analysts advocating an *a priori* position is to choose a tendentious base case that makes their position appear highly attractive.

### 3.4. Program a Change-From-Base cell for Performance Measures

Analytical results are most usefully reported in comparison to the base case. The statement "profit will be $10 million" is less insightful than "profit will increase by $1 million from a base of $9 million". Therefore, the analyst should program a "change-from-base" cell for each performance measure. In a spreadsheet, use Edit\Paste Special…\Values\ to place the performance measure base value in a cell as a "benchmark". In a nearby cell program the **change-from-base** which is the current performance measure minus the (benchmark) base value. When inputs are set to the base case, the change-from-base will be zero. When inputs are not at the base case, the change-from-base will report the difference between the current situation and the base case. See Figure 2.

|   | A | B | C | D | E | F | G | H | I |
|---|---|---|---|---|---|---|---|---|---|
| 8 |   | MODEL |   |   |   |   |   |   |   |
| 9 |   |   | Asralide | Belex | Coroflux | Dalisyn | Total |   |   |
| 10 |   | Sales Revenue | $ 25.84 | $ 50.72 | $ 20.44 | $ 10.40 | $ 107.40 | Total Net Profit |   |
| 11 |   | Contribution | $ 16.02 | $ 37.03 | $ 13.90 | $ 5.62 | $ 72.56 | Value of | Change from |
| 12 |   | Sales Rep Expense | $ 6.00 | $ 12.00 | $ 7.50 | $ 4.50 | $ 30.00 | Base Case | Base Case |
| 13 |   | Net Profit | $ 10.02 | $ 25.03 | $ 6.40 | $ 1.12 | $ 42.56 | $ 38.33 | $ 4.23 |

Figure 2: Change-from-base. Cell H13 contains the "benchmark" base case value for Total Net Profit. Cell I13 contains **=G13-H13**. The current value of Total Net Profit is $4.23 million higher than the base case value.

### 3.5. Record Base Case and Base Values

It is not uncommon in the throes of analysis to overwrite the values of the base case inputs. Thus, it is important to store a copy of the base case input values in a safe place, along with a copy of the base case output values for reference. In a spreadsheet with a single worksheet, this is easily done by setting the inputs to the base case, copying the model, going to a new "base values" worksheet and performing Edit\Paste Special…\Values\ followed by Edit\Paste Special…\Formats\. The base values worksheet should be set to "formula view" to emphasize that it contains only values and be protected so it cannot be inadvertently overwritten. Should the base case values need to be retrieved, copy the input module of the base values worksheet and paste it onto the input module of the spreadsheet. (Because the formats were copied, a simple copy-paste operation suffices). The other base values are a useful reference to confirm that output values have not changed.

## 4. SENSITIVITY ANALYSIS

**Sensitivity analysis** is systematically varying one or more model inputs and recording the impact on one or more model outputs. More formally, sensitivity analysis is when the analyst designates one or more model inputs to be parameters, varies the parameter values while holding all other inputs at their base values, and records the resulting model outputs. The techniques vary depending on the number of inputs being changed.

### 4.1. What-If Analysis

The foundation of sensitivity analysis is "what-if analysis". **What-if analysis** is when the analyst changes one number in a model and observes what happens to the outputs. A change-from-base cell is very helpful. What-if analysis is the simplest analytical technique. All other analytical techniques are an organized series of what-if analyses.

### 4.2. One-Parameter Sensitivity Analysis

**One-parameter sensitivity analysis** is when the analyst designates a single input as a parameter and systematically varies it through a range of values while recording corresponding output value(s). All other inputs remain at their base values. The analytical result is a column of parameter values accompanied by one or more columns of output values.

One can do this manually: change the parameter value, run the model (in a spreadsheet, this is recalculation) and write down the model output; repeat until all desired parameter values have been run. One can automate one-parameter sensitivity analysis in Excel using Data\Table…\ as shown in Figure 3 for a single output; it is easily extended to handle multiple outputs. The results of a one-parameter sensitivity analysis can usefully be presented as a scatter chart.

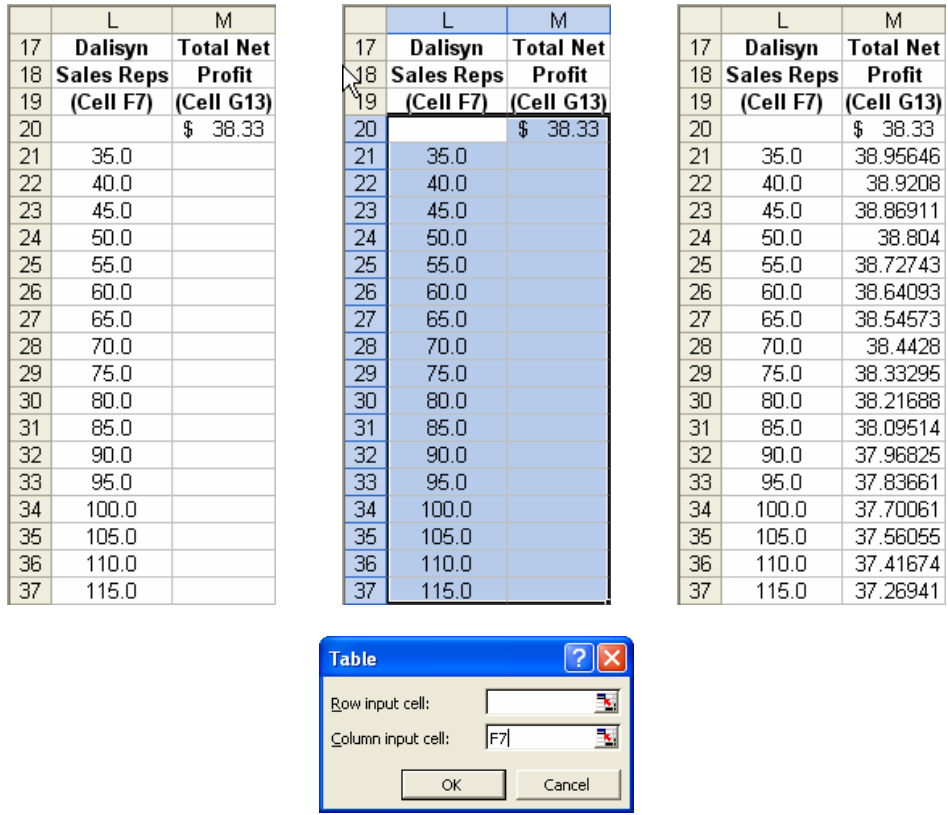

Figure 3: Automating One-Parameter Sensitivity Analysis with a single output using Data\Table…\. The left spreadsheet shows the setup; notice the structure and documentation. The middle spreadsheet with cells L20:M37

selected is ready to perform Data\Table…\. The right spreadsheet shows the results. The filled-out Table dialog box is shown at the bottom.

### 4.3. Comparing Sensitivities: Tornado Chart

A model output is said to be **sensitive** to a model input if that output changes significantly in response to changes in the model input. A model output is **insensitive** to a model input if that output does not change, or changes only slightly in response to changes in the model input. Analysts (and executives) need to know which inputs are sensitive and thus require managerial attention and which are insensitive and thus can be ignored.

A powerful tool for comparing the sensitivities of model inputs is called the **tornado chart** which summarizes one-parameter sensitivity analyses on every model input. I have seen tornado charts provided to the top management of multi-national corporations. The analyst uses managerial judgment to provide a low input value and high input value for each model input. For each input, the resulting pair of output values are charted as a horizontal bar. The length of the bar is a measure of sensitivity of the input. The bars for all inputs are sorted by length and presented with the longest on top and shortest at the bottom. The base output value is shown as a vertical line. See Figure 4.

These calculations can be performed manually and charted. Charting is surprisingly tricky using Excel's chart wizard, so many analysts use a "template" approach where they copy an old tornado chart and change the numbers. I recommend using an add-in.

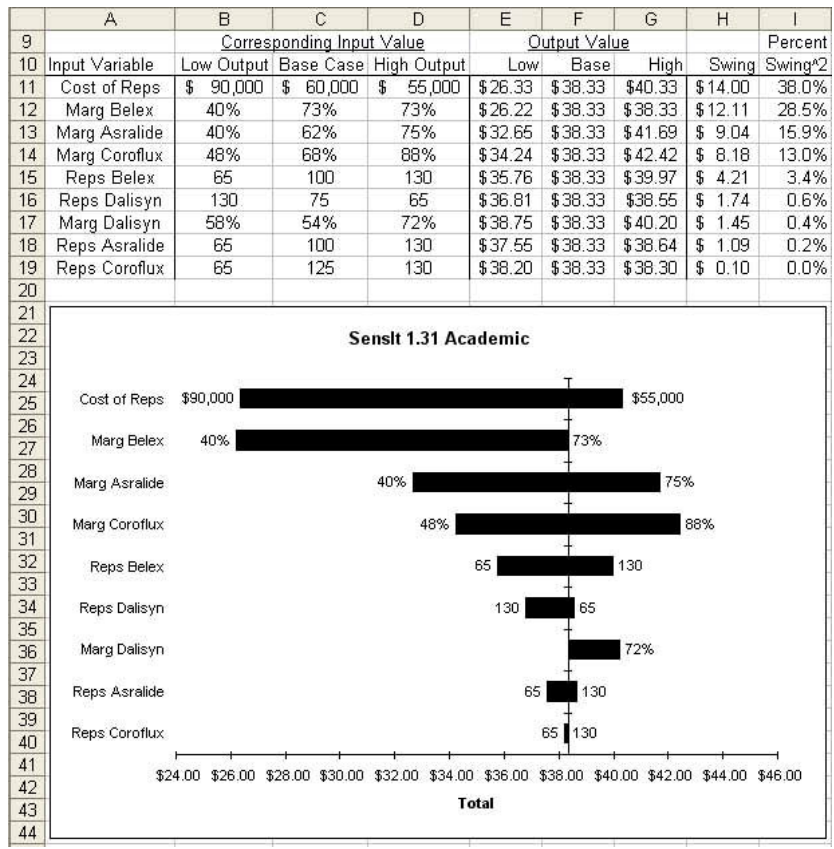

Figure 4: Tornado Chart. The tornado chart is shown at the bottom and the source data for the chart is shown at the top. This example was made using SensIt.

### 4.4. Two-Parameter Sensitivity Analysis

**Two-parameter sensitivity analysis** is when the analyst parameterizes two inputs simultaneously and systematically varies them through a range of values while recording corresponding output value(s). All other inputs remain at their base values. The analytical result is a table of output values with parameter values along the top row and left column.

One can do this manually: change the two parameter values, run the model (in a spreadsheet, this is recalculation) and write down the model output; repeat this process until all desired parameter value pairs have been run. One can automate two-parameter sensitivity analysis in Excel using Data\Table…\ as shown in Figure 5.

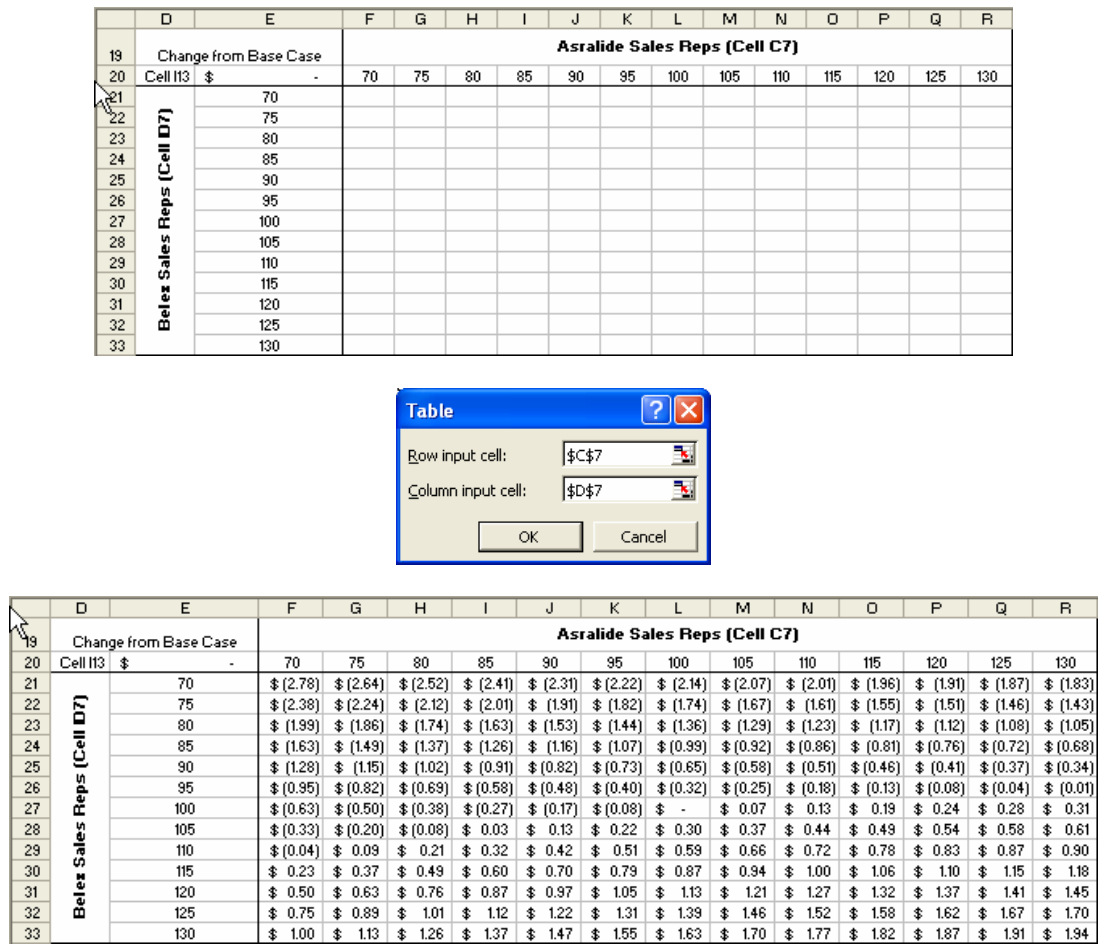

Figure 5: Automating Two-Parameter Sensitivity Analysis using Data\Table…\. The top spreadsheet shows the setup; notice the structure and documentation; cell E20 contains **=I13**. To create the table, select cells E20:R33 and perform Data\Table…\; the filled-out Table dialog box is shown in the middle. The bottom spreadsheet shows the results.

### 4.5. Multi-Parameter Sensitivity Analysis: Scenario Analysis

The tabular approach for sensitivity analysis for varying one or two parameters does not scale up effectively because it is difficult to write a table of numbers with more than two dimensions. When three or more inputs are to be varied the analyst uses scenario analysis. A **scenario** is a set of pre-defined input values, usually with a descriptive name. **Scenario analysis** runs the scenarios through the model and records desired outputs.

Any technology for scenario analysis should satisfy several properties. It should be *transparent* meaning that scenario input values are clearly visible. It should be *flexible* meaning that it is easy to modify the scenario input values. It should be *dynamic* (or "live") meaning that the outputs automatically update should the model be changed or an input not included in the scenarios be modified. It should be *verifiable* meaning that the outputs can be checked for accuracy. Finally a tool for scenario analysis should be *efficient* meaning that an analyst can employ it quickly once the managerial work of defining the scenarios has been completed.

Excel has a "scenario manager" tool (Tools\Scenarios…\) that does not satisfy any of these properties. (The scenario manager and the importance of scenario analysis are discussed in Markham and Palocsay 2006.) Scenario input values are buried in a series of dialog boxes limiting transparency and flexibility. The output values are static (or "dead"). The outputs are not verifiable because there are no cell formulas to trace. It is not efficient because it is cumbersome and slow to use.

The spreadsheet analyst is better served by building what I call a **scenario tool**, shown in Figure 6. The scenario tool is a programming technique that can be performed quickly (on the order of one or two minutes) on any well-designed model where the scenarios have been defined. It uses the INDEX function to grab the specified scenario, which is then echoed to the input cells of the spreadsheet model.

|   | A | B | C | D | E | F | G | H | I |
|---|---|---|---|---|---|---|---|---|---|
| 2 | **INPUTS** | | | | | | | | |
| 3 | | Sales Rep Cost | 60,000 | | | | **Selected Scenario** | | |
| 4 | | | | | | | Number | Name | |
| 5 | | | Asralide | Belex | Coroflux | Dalisyn | 3 | High Margins | |
| 6 | | Contribution Margin | 70% | 88% | 82% | 69% | | | |
| 7 | | Num. Sales Reps | 100.0 | 100.0 | 125.0 | 75.0 | | | |
| 8 | **MODEL** | | | | | | | | |
| 9 | | | Asralide | Belex | Coroflux | Dalisyn | Total | | |
| 10 | | Sales Revenue | $ 25.84 | $ 36.71 | $ 20.44 | $ 10.40 | $ 93.39 | Total Net Profit | |
| 11 | | Contribution | $ 18.09 | $ 32.30 | $ 16.76 | $ 7.18 | $ 74.33 | Value of | Change from |
| 12 | | Sales Rep Expense | $ 6.00 | $ 6.00 | $ 7.50 | $ 4.50 | $ 24.00 | Base Case | Base Case |
| 13 | | Net Profit | $ 12.09 | $ 26.30 | $ 9.26 | $ 2.68 | $ 50.33 | $ 38.33 | $ 12.00 |
| 14 | | | | | | | | | |
| 15 | **CONVERSION of $ to $million** | | | | | | | | |
| 16 | | Sales Rep Cost | $ 0.060 | | | | | | |
| 17 | | | | | | | | | |
| 18 | | | | | | | | | |
| 19 | | | | | | | | | |
| 20 | | | | | | | | | |
| 21 | | | | | | | | | |
| 22 | | | | | | | | | |
| 23 | **SCENARIOS** | | Scenario | Scenario | Sales Rep | Asralide | Belex | Coroflux | Dalisyn |
| 24 | | | Number | Name | Cost | Cont. Marg. | Cont. Marg. | Cont. Marg. | Cont. Marg. |
| 25 | | Values in Use | 3 | High Margins | 60,000 | 70% | 88% | 82% | 69% |
| 26 | | | | | | | | | |
| 27 | | | 1 | Base Case | 60,000 | 62% | 73% | 68% | 54% |
| 28 | | | 2 | Low Margins | 60,000 | 52% | 58% | 50% | 49% |
| 29 | | | 3 | High Margins | 60,000 | 70% | 88% | 82% | 69% |
| 30 | | Scenario Table | 4 | Low Rep Cost | 50,000 | 62% | 73% | 68% | 54% |
| 31 | | | 5 | High Rep Cost | 75,000 | 62% | 73% | 68% | 54% |
| 32 | | | 6 | Pessimistic | 75,000 | 52% | 58% | 50% | 49% |
| 33 | | | 7 | Optimistic | 50,000 | 70% | 88% | 82% | 69% |

Figure 6: Scenario Tool. The scenarios are pre-defined in rows 27-33. Cell G5 contains the desired scenario number. Row 25 uses this scenario number to "grab" the data for that scenario; cell C25 echoes cell G5 using **=G5**; cell D25 contains **=INDEX(D27:D33,$C25)**. Cell D25 is copied into cells E25:I25. The grey input cells in rows 3-6 echo values in row 25; for example cell C3 contains **=E25**.

The analyst can easily generate a summary report by performing a one-parameter sensitivity analysis on the scenario number, as shown in Figure 7. This table is "live" in the sense that any changes to the decision variables in row 7 are instantly reflected in the table.

|    | C | D | E |
|----|---|---|---|
| 35 | Scenario | Scenario | Total |
| 36 | Number | Name | Net Profit |
| 37 | (cell C3) | (cell C8) | (cell G18) |
| 38 |   | Pessimistic | $ 20.04 |
| 39 | 1 | Base Case | $ 38.33 |
| 40 | 2 | Low Margins | $ 26.04 |
| 41 | 3 | High Margins | $ 50.33 |
| 42 | 4 | Low Rep Cost | $ 42.33 |
| 43 | 5 | High Rep Cost | $ 32.33 |
| 44 | 6 | Pessimistic | $ 20.04 |
| 45 | 7 | Optimistic | $ 54.33 |

Figure 7: Summary Report from the Scenario Tool suitable for inclusion in a report or presentation. This report was created by performing a one-parameter sensitivity analysis on Scenario Number using Data\Table…\

**4.6. Automation of Sensitivity Analysis Using Spreadsheet Add-Ins**

There are several add-ins for performing elements of sensitivity analysis. ([Grossman 2004] provides a listing of add-ins; it will be updated in summer 2008.) None of the add-ins is fully satisfactory.

SensIt [SensIt 2008] is professionally-engineered to work across many versions of Excel and operating system and has a PDF user manual. It currently provides one-parameter sensitivity analysis with a single output and tornado chart. Further development is underway. The Sensitivity ToolKit [Sensitivity ToolKit 2007] provides one-parameter sensitivity analysis with multiple outputs, two-parameter sensitivity analysis, and tornado charts along with Solver sensitivity and Crystal Ball sensitivity. It is an educational tool built in an academic environment. The user is required to adjust Excel security settings to "trust" Visual Basic Project.

Crystal Ball has a tornado chart tool that is limited to computing fixed percentage deviations from base; this is conceptually undesirable and is meaningless for base values of zero or small magnitude. Spreadsheet Detective has a simple sensitivity analysis capability that varies inputs by fixed percentage deviations from base.

**5. BACKSOLVING OR GOAL-SEEKING**

**Backsolving** is discovering a value of a model input that produces a specified model output. Common business examples of backsolving include breakeven analysis (how many units to sell to cover fixed costs) and internal rate of return (what discount rate causes net present value to be zero).

Backsolving can be done by manual search or by performing a one-parameter sensitivity analysis on the input to discover which input value(s) yield the specified output; a chart is helpful. In Excel backsolving is automated with Tools\Goal Seek…\. As shown in Figure 8, the analyst indicates the output cell ("set cell"), the specified value of the output cell ("to value"), and the input cell to be varied "By changing cell".

The Goal Seek tool is not without problems. There can be multiple solutions to a backsolving problem (for example, the well-known problem of multiple internal rates of return). Goal Seek cannot detect this. Experimenting with extreme starting values can help identify if there are multiple solutions.

Goal Seek executes a search algorithm that terminates when it is "close enough" to the specified value as determined by the "Maximum Change" field in Excel options. If it is not sufficiently close, the analyst might need to reduce the Excel option "Maximum Change".

Goal Seek can stop and report that it "may not have found a solution" as shown in Figure 8. (It is important that the analyst actually read the Status box!) This can occur because there is no value of the input cell that will cause the output cell to take on the specified value. In older versions of Excel this can occur because Goal Seek stopped with too few iterations. In this case the analyst needs to increase the Excel option "Maximum Iterations".

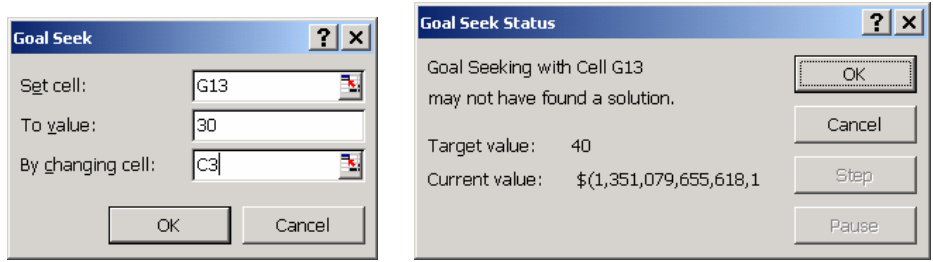

Figure 8: The Goal Seek Dialog Box is on the left. On the right is the Goal Seek Status box when there is no solution or the Goal Seek tool stopped too soon.

## 6. OPTIMIZATION

Optimization is any technique that discovers the "best" values for decision variables. Optimization can be performed using sensitivity analysis for models with only one or two decision variables. For more decision variables it requires an add-in such as the Excel Solver. More powerful solvers are available from Frontline Systems (Premium Solver product line, www.solver.com) and Lindo Systems (What's Best! product line, www.lindo.com). These add-ins implement sophisticated mathematical algorithms that can guarantee that the resulting decision variable values achieve the highest (or lowest) possible value of the performance measure, taking into account business rules called "constraints". Crystal Ball has a limited optimization capability. Frontline recently rewrote the Excel calculation engine for Microsoft and has developed "Psi Technology" that allows instant conversion of most spreadsheet code into C++ code which runs faster and allows for more advanced mathematical analysis.

It is important to distinguish between the model—which is a machine that converts inputs to outputs—and the optimization algorithm which is a "black box" that proposes decision variable values and asserts that those values have certain mathematical properties.

## 7. RISK ANALYSIS: MONTE CARLO SIMULATION

Many decisions need to be made in the face of uncertainty. In a model, uncertainty boils down to a data input whose value we don't yet know. Sensitivity analysis can teach the analyst about the impact of different possible values of uncertain data inputs. Monte Carlo simulation combines the impact of different possible input values with the likelihood of those values occurring. It works by replacing a fixed input value with a probability distribution. By randomly sampling each probabilistic input the analyst can generate any number of equally likely possible future states. For each of these states, the spreadsheet calculates the corresponding output values, and these multiple output values in aggregate can be thought of as an estimate of the probability distribution of the output. The analyst can examine this output distribution for its average, median, percentiles, standard deviation, probability of extreme events and so forth.

When using a spreadsheet, Monte Carlo simulation can with difficulty be performed without the use of an add-ins, but this is painful and only arguably appropriate even for educational use. For any practical purpose it is essential to use an add-in. The well-established high-end commercial add-ins are Crystal Ball (by Oracle) and @Risk (by Palisade Corporation). A new high-end commercial product called Risk Solver from Frontline Systems has many innovations and could be a powerful competitor. Less-expensive alternatives include RiskSim from Decision Toolworks and XLSim from analycorp.com.

## 8. CONCLUSIONS, FURTHER RESEARCH, AND SOFTWARE DEVELOPMENT NEEDS

This paper presents a hierarchy of analytical techniques that apply to any model. It discusses how to implement each technique efficiently in Microsoft Excel, and discusses available add-ins for automating these techniques.

We know very little about the use of spreadsheet analytic techniques in practice. We would benefit from empirical research to understand the number and scope of spreadsheets used for analysis (as opposed to some other purpose), the prevalence of various analytical techniques, and the ability of people to name and discuss the analytical techniques they employ. [Baker et al 2006] provide insight into use of tools such as Data Table, Solver, and Goal Seek.

Analysts would benefit from an improved sensitivity analysis add-in. An ideal sensitivity analysis add-in would have the following attributes: 1) high-quality software engineering, 2) function on multiple operating systems, 3) a user manual, 4) technical support, and 5) the resources of a software company. An ideal sensitivity analysis add-in would have at a minimum the following functionality: 1) one-parameter sensitivity analysis with multiple outputs plus chart, 2) two-parameter sensitivity analysis with multiple outputs (one table per output), 3) tornado chart that accepts user-specified values for the low and high value of each input (not fixed percentage change).

The product that is closest to this list of desired features is [SensIt 2008] by Decision Toolworks, which has high-quality software engineering and runs on multiple platforms. Technical support is available. However it is a one-man shop with limited resources. SensIt has one-parameter sensitivity analysis with single output (multiple outputs is planned) plus chart, no capability for two-parameter sensitivity analysis (this is also planned), and an excellent tornado chart capability.

Analysts would benefit from an "optimization sensitivity" add-in that varies one or more parameters and runs Solver for each. The existing add-ins are [Sensitivity ToolKit 2007] and [SolverTable 2007]. These provide effective proof-of-concept and can be suitable for educational application, but they are not robust tools for the analytical professional. A satisfactory product would have the same attributes as discussed previously for sensitivity analysis add-ins: 1) high-quality software engineering, 2) function on multiple operating systems, 3) a user manual, 4) technical support, and 5) the resources of a software company.

## ACKNOWLEDGEMENTS

I am grateful to Stephen Powell at the Tuck School of Business for sharing with me the hierarchy of skills and establishing the outline of ideas that are explored in this paper.

## DISCLOSURES

Decision Toolworks is owned by Mike Middleton, my colleague at the University of San Francisco. I was an alpha-tester and beta-tester for Sensitivity ToolKit. I have no financial relationship with Decision Toolworks nor Sensitivity ToolKit.